\documentclass[twocolumn,twoside,slac_two]{revtex4}
\usepackage{graphicx}
\usepackage{fancyhdr}
\begin{document}

\title{
The Bologna Complete Sample: radio and gamma-ray data.}

\author{E. Liuzzo, G. Giovannini}
\affiliation{Istituto di Radioastronomia, Via P. Gobetti 101, 40129 Bologna (Italy)}
\affiliation{Dipartimento di Astronomia, Universit\`a di Bologna, Via Ranzani 1, 40127 Bologna (Italy)}

\author{M. Giroletti}
\affiliation{Istituto di Radioastronomia, Via P. Gobetti 101, 40129 Bologna (Italy)}

\begin{abstract}
To study a statistical properties of different classes of radio sources, we defined and observed the Bologna Complete Sample (BCS) which is unbiased with respect to the orientation of the nuclear relativistic jet being selected from low-frequency samples. The BCS is a complete sample of 94 nearby (z$<$0.1) radio galaxies that are well studied targets with literature kiloparsec data. For all of them, we collected parsec scale
information asking new VLBI observations. Statistical results on their properties in radio band are presented. From the estimates of the Doppler factor and viewing angles, we discuss the connection with the available gamma-ray data.
Finally, we show how future observations with Fermi could reveal new important detections of some of the BCS sources.
\end{abstract}

\maketitle

\thispagestyle{fancy}

\section{INTRODUCTION}
The statistical study of the parsec scale properties of different classes of radio galaxies is crucial to obtain information on the nature of their central engine. To this aim, it is important to define and observe a sample that is free from selection effects looking at radiogalaxies in low radio frequencies survey. In fact, sources in low-frequency samples are dominated by their extended and unbeamed (isotropic) emission, rather than the beamed compact emission that dominates in high-frequency studies. Low frequency surveys are therefore unbiased with respect to the orientation of the nuclear relativistic jet. With this purpose in mind, we initiated a project to investigate a complete sample of radio galaxies selected from the B2 Catalogue of Radio Sources and the Third Cambridge Revised Catalogue (3CR) (\cite{gio01, gio90}), with no selection constraint on the nuclear properties. We named this sample ``the Bologna Complete Sample'' (BCS).

In the original sample, 95 radio sources from the B2 and 3CR catalogues were present, but because of the rejection of one source (\cite{liu09, gio05}), we redefined the complete sample to be 94 sources. We selected the sources to be stronger than a flux density limit of 0.25 Jy at 408 MHz for the B2 sources and greater than 10 Jy at 178 MHz for the 3CR sources (\cite{fe84}). We also applied the following criteria: 1) declination $>$ $10^{\circ}$; 2)Galactic latitude $\vert b\vert>15^{\circ}$; 3)redshift z$<$0.1. 
As our main goals was to properly study the central engine of this sample, we asked and obtained high resolution observations with VLBI technique (\cite{liu09, gio05, gio01}) for all sources not yet analysed at this spatial scale.

The Large Area Telescope (LAT) on board Fermi, with its large field of
view and unprecedented sensitivity, is now putting us in the condition of a better
understanding of the extragalactic gamma-ray source population. In anticipation of
the launch of Fermi, large projects in the radio band have been undertaken (e.g. \cite{fu07, li09}). The results of these projects can now be exploited to gain insights into
the radio properties of this population and into the relation between radio and
gamma-ray properties. With these latter purposes, we decided to compare Fermi and results in radio band for our complete sample of nearby radiogalaxies.
In the following, we describe our study and we report its most important conclusions.

\section{RADIO DATA}

Up to now, at parsec scale and in radio band, we analysed 76 sources. Our main results are:
\begin{itemize}
\item The detection rate is high: only 3 sources out of 76 (4$\%$) have not been detected, even though we observed sources with an arcsecond core flux density as low as 5 mJy at 5 GHz. This result confirms the presence of compact radio nuclei at the center of radio galaxies.
\item As expected in sources with relativistic parsec-scale jets, the one-sided jet morphology is the predominant structure present in our VLBI images, however $22\%$ of the observed sources show evidence of a two-sided structure. This result is in agreement with a random orientation and a high jet velocity ( $\beta \sim 0.9$).
\item We find two sources (4C26.42, \cite{liu09b} and 3C 310, Fig.~\ref{1346}) with a Z-shaped structure on the parsec-scale suggesting the presence of low velocity jets in these peculiar radio sources.
\newcommand{\imsize}{1\columnwidth}
\begin{figure*} [th]
\begin{center}
\includegraphics[width=110mm]{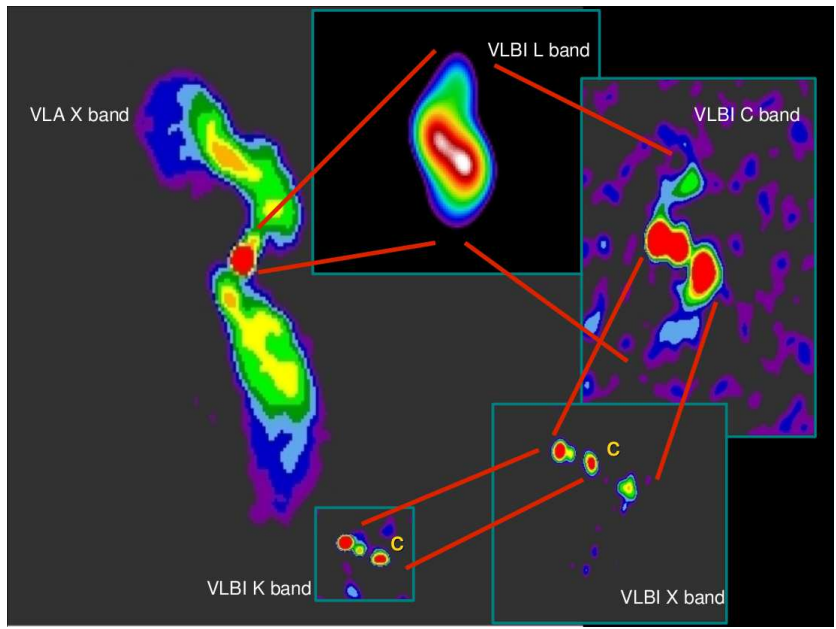}
\end{center}
\caption{Clockwise from left to right, zooming from kiloparsec to mas scale radiostructure of 4C 26.42 (\cite{liu09b}): color maps of VLA X band, VLBI L band, VLBI C band, VLBI X band and VLBI K band data. (C) indicates the core component.  }
\label{1346}
\end{figure*}
\item In 8 sources, the low core dominance suggests that the nuclear activity is now in a low activity state. The dominance of the extended emission implies a greater activity of the core in the past. However in these sources a parsec-scale core and even jets are present. In this scenario the nuclear activity may be in a low or high state but is not completely quiescent. This result is in agreement with the evidence that a few sources show evidence of a recurring or re-starting activity. This point can be better addressed when observations are available for the full sample so that we can discuss the time-scale of the recurring activity.
\item In most cases, the parsec and the kiloparsec scale jet structures are aligned and the main jet is always on the same side with respect to the nuclear emission. This confirms the idea that the large bends present in some BL Lacs sources are amplified by the small jet orientation angle with respect to the line-of-sight.
\item In 62$\%$ of the sources, there is good agreement between the arcsecond-scale and the VLBI correlated flux density. For the other 38$\%$ of the sources, at the milliarcsecond scale more than 30$\%$ of the arcsecond core flux density is missing. This suggests the presence of variability, or of a significant sub-kiloparsec-scale structure, which will be better investigated with the EVLA at high frequency or with the e-MERLIN array.
\end{itemize}

\newcommand{\imsize}{1\columnwidth}
\begin{figure*}[th!]
\begin{center}
\includegraphics[width=160mm]{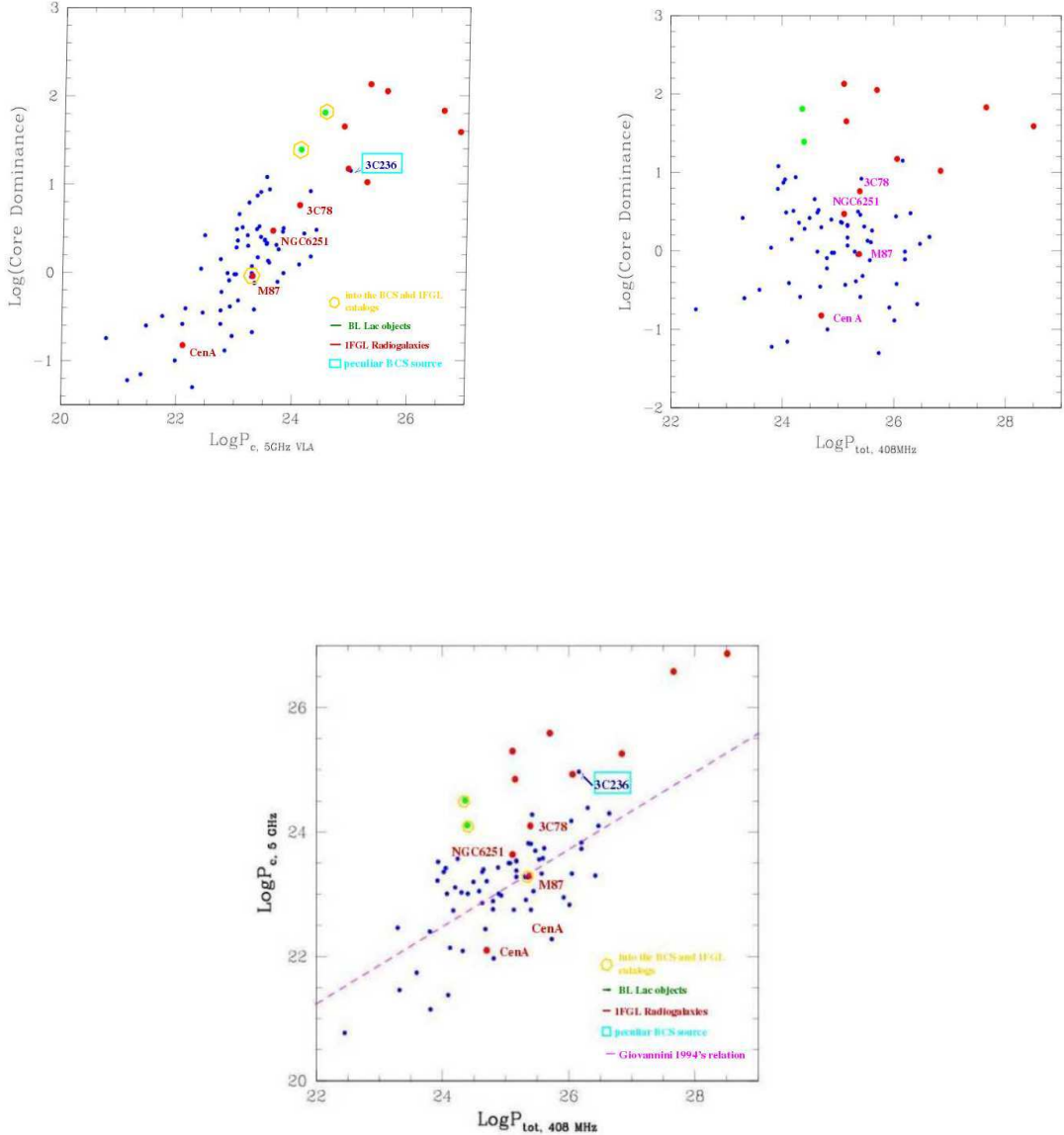}
\end{center}
\caption{{\it Top left}: LogP$_{core}$ versus CD for BCS sources (blue points) plus the RGs in the 1FGL (red points), where P$_{core}$ is the total arcsecond power of the core at 5 GHz. Yellow circles indicate BCS that appears also in the Fermi catalog, green points correspond to BL Lacs BCS sources, azure square identify peculiar BCS radiogalaxy. {\it Top right}: We show Log P$_{tot}$ versus CD where P$_{tot}$ is the total power at 408 MHz. {\it Bottom}: We plotted LogP$_{core}$ versus LogP$_{tot}$ . }
\label{plot}
\end{figure*}

\section{GAMMA-RAY DATA.}

We search in the 1 Year Fermi catalog (1FGL, \ref{fermi1}) available information of the gamma-ray emission for all the BCS sources. We found that: 

\begin{itemize}
\item Among the BCS,there are
3 sources in 1FGL catalog :2 BL Lacs (Mkn 421 and Mkn 501) plus M87;
\item Thanks to radio data, we also estimated the core dominance (CD). According to \cite{gi94} and references therein, a correlation is present between the core and total radio power and we can use the core dominance to estimate the jet velocity and orientation. We calculated the core dominance defined as the ratio between the observed and the estimated core radio power according to the relation given in \cite{gio94} (named here GC). As expected, among the 1FGL Radiogalaxies (RG), the majority of sources are above the Giovannini 1994’s correlation (GC) having high CD which indicates small angle ($\theta$) of view;
\item M87 and Cen A are peculiar objects being below the GC while 3C 236 is interesting being not in the 1FGL Catalog
(see Notes on sources below);
\item From kiloparsec VLA measurements, we derived the total power at 5 GHz of the core ($P_{core}$) and the total power of the source at 408 MHz (P$_{core}$). In the P$_{core}$ - CD plane (top left panel in Fig.~\ref{plot}), 1FGL RGs lie in a well defined region with high CD and high P$_{core}$.
\item P$_{core}$ and P$_{tot}$ do not show any evident correlation (top right panel in Fig.~\ref{plot}).
\item In the P$_{tot}$ $-$ CD plane (bottom panel in Fig.~\ref{plot}), the 1FGL RGs spread over P$_{tot}$, meaning independence of Fermi emission from P$_{tot}$, but they are segregate at high CD values that suggests the Independence of the Fermi emission from P$_{tot}$, but the presence of a relation with P$_{core}$, due to beaming effects plus also intrinsically high P$_{core}$ .
\end{itemize}

\subsection{Notes on individual sources.}
Three objects have peculiar behavior and need notes:
\begin{itemize}
\item {\bf M87}  has small CD and large $\theta$ but it is detected by Fermi as a
consequence of its proximity (z=0.04).
\item {\bf 3C 236} has P$_{core}$, P$_{tot}$ and CD similar to the 1~FGL RGs but it
is not a Fermi source. In this case, the high CD is due to the
restarted activities. It is
like a young source. It could be detected by Fermi in the future.
(\cite{or11}).
\item {\bf Cen A}, despite its small CD, is a 1FGL sources as the Fermi
emission is not only nuclear but it comes also from lobes.
\end{itemize}

\section{CONCLUSION.}

To study a statistical properties of different classes of radio sources, we selected a complete sample of nearby radio galaxies that is free from selection effects, the BCS sample. To properly describe their parsec scale emission, VLBI observations are obtained for all these sources and up to now 76 of them are analysed by us. The one-sided jet morphology is the predominant structure present in our VLBI images, in agreement with a random orientation and a high jet velocity of the source.  However some cases are peculiar showing the presence of low velocity jets or restarted activity. Since these sources have a very faint nuclear emission in the radio band, the remaining 18/94 sources are observed with very sensitive EVN and VLBA observations at 1.6 GHz employing large bandwidths, long integration times, and phase referencing. The work on these data is still in progress and our results on them will be soon published (\cite{liu11}). At the same time, to investigate the gamma-ray emission of the BCS sample, we search in the 1 FGL information for all of them. Three BCS radiogalaxies are 1FGL sources: 2 BL Lacs and M 87. We compared our sample with the 1 FGL RGs. We found that for 1 FGL RGs, the P$_{core}$ correlates with P$_{tot}$ and with the CD, with typically high values of CD, P$_{core}$ and P$_{tot}$. Two sources are detected by Fermi despite their small CD: M87 because of its proximity, and CenA for which the gamma-ray emission comes from the lobes. Thanks to the unprecedented sensitivity and positionally accuracy of Fermi, we expect that other BCS sources with low CD will be detected in the next future, allowing us to better investigate the correlation between gamma-ray and radio emission in nearby radiogalaxies.\\

\begin{acknowledgments}
We thank the organizers of a very interesting meeting.
This work was supported by contributions of European Union, Valle D’Aosta Region and the Italian Minister for Work and Welfare. 
\end{acknowledgments}

\end{document}